\documentstyle[12pt]{article}

\begin{document}

\hfill{NCKU-HEP-98-03}\par

\vfill
\centerline{\large{{PQCD study of the $B_{\rm SL}$ and $n_c$ controversy
in inclusive $B$ decays}}}\par
\vskip 1.0cm 
\centerline{Chia-Hung V. Chang}
\vskip 0.3cm
\centerline{Center for Theoretical Sciences, Hsin-Chu, Taiwan, R.O.C.}
\vskip 0.3cm
\centerline{Darwin Chang, We-Fu Chang}
\vskip 0.3cm
\centerline{Department of Physics, National Tsing-Hua 
University, Hsin-Chu, Taiwan, R.O.C.}
\vskip 0.3cm
\centerline{Hsiang-nan Li}
\vskip 0.3cm
\centerline{Department of Physics, National Cheng-Kung
University, Tainan, Taiwan, R.O.C.}
\vskip 0.3cm
\centerline{Hoi-Lai Yu}
\vskip 0.3cm
\centerline{Institute of Physics, Academia Sinica, Taipei, Taiwan, 
R.O.C.} 
\vskip 1.0cm
\centerline{\today }
\vskip 2.0 cm
\centerline{\bf Abstract}

We calculate the semileptonic branching ratio and the charm counting of 
inclusive $B$ meson decays using the perturbative QCD formalism.
For the nonleptonic decays, we employ the modified
factorization theorem, in which Wilson coefficients evolve with the
characteristic scales of the decay modes. It is found that
the decay rate of the single-charm mode $b\to c{\bar u}d$ is enhanced, and 
a lower $B_{SL}$ is obtained without increasing $n_c$.  We predict a
larger $b\to c \tau \bar{\nu}$ branching ratio compared to that from the
conventional heavy-quark-effective-theory based operator product expansion. 
Our result of the $B$ meson lifetime is also consistent with the data.
\vfill
\newpage

With rapid theoretical and experimental developments, the heavy flavor
physics has reached the level of accuracy that the data can be used to 
test the standard model. Especially, the various quantities related to the 
heavy meson decays have been calculated and compared with experimental data. 
In spite of many successes, a discrepancy exists consistently between the
measurements and the theoretical predictions of the inclusive semileptonic
branching ratio $B_{\rm SL}$ and the charm counting $n_c$ of the $B$ meson
decays: The latest results from CLEO, measured at the $\Upsilon(4s)$
resonance, are $B_{\rm SL}=(10.19 \pm 0.37) \%$ and $n_c=1.12 \pm 0.05$
\cite{CLEO}, while LEP reports, at the $Z$ resonance,
$B_{\rm SL}=(11.12 \pm 0.20) \%$ and $n_c=1.20 \pm 0.07$ \cite{LEP}. The
naive parton model, which coincides with the leading-order
heavy-quark-effective-field-theory (HQEFT) calculation, gives
$B_{\rm SL} \approx 13\%$ \cite{a6}. Conventionally, the inclusive $B$
meson decays are described by a systematic HQEFT based Euclidean space
operator product expansion (OPE) of relevant hadronic matrix elements in
the inverse power of the $b$ quark mass $m_b$.  The assumption of global
and local quark-hadron dualities is employed to justify this approach. The
$1/m_b^2$ corrections (there are no $1/m_b$ corrections) have been
investigated and found to be less than $5\%$ of the leading-order
predictions \cite{neubert}. Hence, it was argued that the values of
$B_{\rm SL} < 12.5\%$ can not be accommodated by QCD theory \cite{af}, and
this discrepancy could be an indication of ``new physics'' beyond the
standard model. For example, an enhanced $b\to s g$ vertex has been
proposed to explain the anomaly \cite{kagan}.

It is quite possible that the assumption of the local duality fails
at the $1/m_b$ level. Recently, there was theoretical indication that
$1/m_b$ corrections might appear in nonleptonic decays \cite{grinstein}.
Altarelli {\it et al.} has observed that if $m_b$ in the phase
space factor was replaced by the $B$ meson mass $M_B$, and by
$M_{\Lambda_b}$ in the $\Lambda_b$ decays, due to the breakdown of the
local duality, the predictions agree well with the data of $B_{\rm SL}$
and of the lifetime ratio \cite{A}. This naive modification increases the
nonleptonic decay rates, and thus reduces $B_{\rm SL}$. On the other hand,
though the nonperturbative corrections are small, it was pointed out
that perturbative corrections may be significant \cite{a6}. Bagan
{\it et al.} have computed the $O(\alpha_s)$ corrections with the effects
of the charm mass included \cite{bagan}, and found that $B_{\rm SL}$ is
indeed lowered with the increase of the $b\to c{\bar c}s$ rate.
However, the predictions are sensitive to the choice of the renormalization
scale $\mu$ \cite{NS}: $B_{\rm SL}=12.0$ for $\mu=m_b$, and
$B_{\rm SL}=10.9$ for $\mu=m_b/2$. Furthermore, an enhanced $b\to c\bar{c}s$
mode, while lowering $B_{\rm SL}$, enlarges $n_c$ to around $1.20$, which,
though consistent with the LEP data, becomes significantly higher than the
CLEO data. This is sometimes called the ``missing charm puzzle''. 

In this paper we shall analyze the inclusive $B$ meson decays based on
perturbative QCD (PQCD) factorization theorems \cite{sterman}. It will be
shown that the experimental data of $B_{SL}$ and $n_c$ can be accommodated
simultaneously within the framework of the standard model, without resort to
new physics. In a series of recent works \cite{LiYu,LiL} we have developed
a PQCD formalism for the inclusive semileptonic decay $B\to X_ue{\bar \nu}$,
whose decay rate is written as the convolution of a hard subamplitude with
the $B$ meson distribution function and the $u$-quark jet function. The
idea is briefly explained as follows: Loop corrections produce infrared
divergences, which signal the sensitivity to long-distance physics.
The soft divergences associated with vanishing
loop momenta survive after summing over all diagrams of the same order of
$\alpha_s$. These divergences are absorbed into the $B$ meson distribution
function, which describes the probability that the $b$ quark carries
some fraction of the $B$ meson momentum. This function has been identified
as the soft function associated with the Fermi motion of the heavy quark
inside the heavy hadron, and as the outcome of the resummation of
nonperturbative power corrections in HQEFT \cite{N}. Near the end points of
the decay spectra, the invariant mass of the $u$ quark vanishes, and the
collinear divergences due to radiative gluons parallel to the $u$ quark
become important. Large double logarithms from the overlap of the soft and
collinear divergences then arise, and are absorbed
into the jet function. The remaining finite piece of the radiative
corrections is grouped into the hard subamplitude.

The hard subamplitude, dominated by short-distance contributions, is
evaluated using perturbation theory. The end-point double logarithms
contained in the jet function are systematically resumed into a Sudakov
factor. It turns out that the Sudakov factor 
suppresses the long-distance contributions, and renders
the perturbative calculation more reliable and self-consistent. The
nonperturbative $B$ meson distribution function is universal, and
determined by other experiment data, such as the spectrum of the inclusive
radiative decay $B\to X_s\gamma$ \cite{btoxgamma}. Note that the analysis
of inclusive heavy meson decays suffers the infrared renormalon ambiguity,
which appears as power corrections starting with $O(1/m_b^2)$. This
ambiguity can be parametrized as an exponential factor associated with the
distribution function, which has also been fixed from the best fit to the
$B\to X_s\gamma$ spectrum \cite{btoxgamma}. The above prescriptions
specify clearly the treatment of each subprocess in the formalism.
The advantage of our approach lies in that it is formulated according
to the meson kinematics, instead of the parton kinematics as used in 
HQEFT. Hence, the replacement of $m_b$ by $M_B$ postulated in \cite{A} is
implemented naturally. The kinematic gap between the parton
picture and the hadron picture is also removed.

We apply the above PQCD factorization theorem to the charmed decays,
\begin{equation}
B(P_{B}) \rightarrow X_{c}+l(p_l)+\bar{\nu}(p_{\nu}),
\end{equation}
which dominate the inclusive semileptonic $B$ meson decays. The momenta are
expressed, in terms of light-cone coordinates, as
\begin{equation}
 P_B=\left(\frac{M_B}{\sqrt{2}},\frac{M_B}{\sqrt{2}},0_{\perp}\right),
 \mbox{\ \ }
     p_l=(p^+_l,p_l^-,0_{\perp}),\mbox{\ \ }
     p_{\nu}=(p_{\nu}^+,p_{\nu}^-,{\bf p}_{\nu \perp}),
\end{equation}
where the minus component $p_l^-$ vanishes for massless leptons. We choose
$E_l$, $q^2$ and $q_0$, or equivalently, the three dimensionless quantities,
\begin{equation} 
x=\frac{2E_l}{M_B}\;,\mbox{\ \ } y=\frac{q^2}{M_B^2}\;,\mbox{\ \ }
y_0=\frac{2q_0}{M_B}\;,
\label{dl}
\end{equation}
as the independent variables, $E_l$ being the lepton energy and
$q \equiv p_l+p_{\nu}$ the momentum of the lepton pair. Their kinematic
ranges are
\begin{eqnarray}
\frac{2m_l}{M_B}\leq&x& \leq 1-\beta+\alpha,
\nonumber\\
\alpha\leq &y& \leq \alpha+(1+\alpha-\beta-x)
\frac{x+\sqrt{x^2-4\alpha}}{2-x-\sqrt{x^2-4\alpha}},
\nonumber\\
x+\frac{2(y-\alpha)}{x+\sqrt{x^2-4\alpha}}\leq &y_0&\leq 1+y-\beta,
\end{eqnarray}
with the constants
\begin{equation}
\alpha \equiv \frac{m^2_l}{M^2_B}, \;\;\;\;
\beta \equiv \frac{M^2_D}{M^2_B}\;,
\end{equation}
and the lepton mass $m_l$ and the $D$ meson mass $M_D$. $M_D$ is introduced
as the minimal invariant mass of the decay product $X_c$. It is easy to
check that the above ranges reduce to those for the decay
$B\to X_u e{\bar \nu}$ as $m_l$, $M_D\to 0$ \cite{LiYu}.

In the parton picture the $B$ meson is composed of a $b$ quark with momentum 
$p_b=P_B-p$, where $p=(p^+,0,{\bf p}_\bot)$ is the momentum carried by other
light degrees of freedom. Since the $c$ quark is massive, there are no 
collinear and thus no large double logarithms associated with it,
differing from the decay $B\to X_u e{\bar \nu}$ \cite{LiYu}. That is, QCD
corrections to the charmed semileptonic decays are weaker. A convenient
factorization scale is the transverse momentum $p_{\perp}$, or its
conjugate variable $b$, of the $b$ quark inside the $B$ meson \cite{LiRe}.
$1/b$ marks the scale separation, below which physics is absorbed into
the $B$ meson distribution function $f$, and above which physics is
absorbed into the hard $b$ quark decay subamplitude $H$. Therefore, we
write the decay rates as,
\begin{eqnarray}
\frac{1}{\Gamma^{(0)}_\ell}\frac{d^3\Gamma}{dxdydy_0}
=\frac{M_B^2}{2}\int^1_{z_{\rm min}}{dz} 
\int_0^\infty db b f(z,b){\tilde J}_c(b)H(z)\;,
\label{asb}
\end{eqnarray}
with $\Gamma^{(0)}_\ell \equiv (G_F^2/16\pi^3)|V_{cb}|^2M^5_B$.
The momentum fraction $z$ is defined as $z \equiv p^+_b/P^+_B=1-p^+/P^+_B$.
$z$ is equal to the maximum 1 as the $b$ quark takes all the $B$ meson 
momentum in the plus direction. The minimum of $z$ is determined by the 
condition $p_c^2 > m_c^2$, $p_c$ and $m_c$ being the $c$ quark momentum
and mass, respectively, which leads to
\begin{equation}
z_{\rm min} =
\frac{\displaystyle\frac{y_0}{2}-y+\frac{m_c^2}{M_B^2}-
\frac{x}{\sqrt{x^2-4\alpha}} \left[-\frac{y_0}{2}+\frac{y}{x}+
\frac{\alpha}{x}\right] }
{\displaystyle 1-\frac{y_0}{2}-\frac{x}{\sqrt{x^2-4\alpha}}
\left[-\frac{y_0}{2}+\frac{y}{x}+\frac{\alpha}{x} \right] }.
\end{equation}
The function ${\tilde J}_c$ denotes the final-state cut, which is in fact
part of the hard subamplitude. We factorize it out in order to compare 
Eq.~(\ref{asb}) with the formulas in the nonleptonic case below. $J_c$ and
$H$ in momentum space are given by
\begin{eqnarray}
J_c&=& \delta \left(p_c^2-m_c^2\right)
\nonumber \\
 &=& \delta\left( M_B^2\left\{z-(1+z)\frac{y_0}{2}+y+
\frac{x(1-z)}{\sqrt{x^2-4\alpha}}
\left[-\frac{y_0}{2}+\frac{y}{x}+\frac{\alpha}{x}\right]
-\frac{m_c^2}{M_B^2}\right \}- {\bf p}^2_{\perp}
-2{\bf p}_{\nu\perp}\cdot{\bf p}_{\perp}\right),
\nonumber \\
& &\label{jc}\\
H &\propto& (p_b \cdot p_{\nu})(p_l \cdot p_c)  
\nonumber\\
&\propto& 
\left((y_0-x)\left\{1-\frac{(1-z)}{2}
\left(1-\frac{x}{\sqrt{x^2-4\alpha}}\right)\right\}
-\frac{(1-z)}{\sqrt{x^2-4\alpha}}
(y-\alpha)\right)\nonumber\\
&&\times \left(\frac{x}{2}\left\{1+z+(1-z)
\frac{\sqrt{x^2-4\alpha}}{x}\right\}
-y-\alpha\right).
\label{hi}
\end{eqnarray}
The universal distribution function determined from the decay spectrum
of $B \rightarrow X_s \gamma$ is written as \cite{btoxgamma},
\begin{equation}
f(z,b)= \frac{0.02647 z(1-z)^2}{[( z - 0.95 )^2 + 0.0034 z]^2} 
\exp( -0.0125 M_B^2 b^2)\;,
\label{bdf}
\end{equation}
where the exponential parametrizes the infrared renormalon contribution
at large $b$.

We then extend our formalsim to the nonleptonic decays
$B\rightarrow X_{c\bar{c}}$ and $B \rightarrow X_{c}$, which are dominated
by the $b$ quark decays $b\to c{\bar c}s$ and $b\to c{\bar u}d$,
respectively. For simplicity, we ignore the pretty small charmless
processes. To simplify the calculation, we route the transverse momentum
$p_\bot$ of the $b$ quark through the $c$ quark jet as in the
semileptonic cases, such that the $c$ quark momentum is the same as before.
For kinematics, we make the correspondence with the ${\bar c}$ (${\bar u}$)
quark carrying the momentum of the massive (light) lepton $\tau$ ($e$ and
$\mu$) and with the $s$ and $d$ quarks carrying the momentum of ${\bar\nu}$.
The dimensionless variables are then defined exactly as Eq.~(\ref{dl}). 
The nonleptonic decays involve more complicated strong interactions
characterized by three scales: the $W$ boson mass $M_W$, the characteristic
scale $t$ of the $B$ meson decay, and the factorization scale $1/b$. The
hard gluon exchanges among the quarks generate logarithms of $M_W$, and the
associated physics is factorized into a {\em harder} function
\cite{LiCh}. Collinear divergences and soft divergences from radiative
corrections exist simultaneously, when the light quarks, such as ${\bar u}$,
$d$, and $s$, become energetic. The resultant double logarithms 
$\ln^2({\bar p}b)$, ${\bar p}$ being the jet momentum defined later, are 
factorized into the corresponding jet functions. This is how the additional 
scales $M_W$ and ${\bar p}$ are introduced
into the nonleptonic decays. There are also soft gluon exchanges among the
quarks, whose effects are negligible in bottom decays \cite{LiTs}.
Similarly, physics with the scale below $1/b$ is absorbed into the
(soft) distribution function.

We employ the three-scale factorization theorem developed in \cite{LiCh},
that embodies both the PQCD factorization theorem stated above and effective
field theory. According to the above explanation, the decay rates are
expressed as the convolution of the harder, hard, jet and soft subprocesses,
\begin{eqnarray}
\frac{\Gamma}{\Gamma_0}=M_B^2\int dxdydy_0 dz
\int \frac{d^2{\bf b}}{(2\pi)^2}
H_r(M_W,\mu)H(z,t,\mu)\prod_i {\tilde J}_i(b)\prod_j {\tilde J}_j(b,p_j,\mu)
{\tilde S}(z,b,\mu)\;,
\label{as}
\end{eqnarray}
where the renormalization scale $\mu$ denotes the inclusion of QCD
corrections. The functions ${\tilde J}_i$, $i=c$ and ${\bar c}$ for the
$b \rightarrow c\bar{c}s$ mode and $i=c$ for the $b \rightarrow c\bar{u}d$
mode, do not contain double logarithms, and represent the final-state cuts.
The light-quark jet functions ${\tilde J}_j$, $j=s$ for the
$b \rightarrow c\bar{c}s$ mode and $j={\bar u}$, $d$ for the
$b \rightarrow c\bar{u}d$ mode, contain double logarithm. These double
logarithms are resummed into a Sudakov factor \cite{LiYu,LiL}, 
\begin{equation}
{\tilde J}_j(b,p_j,\mu)=\exp \lbrack -2 s({\bar p}_j,b)\rbrack
{\tilde J}_j(b,\mu)\;,
\label{su}
\end{equation}
where the upper bound of the Sudakov evolution is chosen as the sum of the
longitudinal components of $p_j$,
\begin{equation}
{\bar p}_j\equiv (p_j^++p_j^-)\;.
\end{equation}
We refer to \cite{BS} for the expression of the exponent $s$. In the
numerical analysis below $\exp(-s)$ is set to unity as ${\bar p}_j< 1/b$.

The single logarithms in the various subprocesses are summed using
the RG equations \cite{LiYu,LiCh}:
\begin{eqnarray}
{\tilde J}_j(b,\mu)&=&\exp\left[ -\int^{\mu}_{1/b}
\frac{d\bar\mu}{\bar\mu}\gamma_{J_j}(\alpha_s(\bar\mu))\right]
\tilde J_j(b)\;,
\nonumber\\
{\tilde S}(z,b,\mu)&=&\exp\left[ -\int^{\mu}_{1/b}\frac{d\bar\mu}{\bar\mu}
\gamma_S(\alpha_s(\mu))\right] f(z,b)\;,
\nonumber\\
H_r(M_W,\mu)&=&\exp\left[ -\int^{\mu}_{M_W}\frac{d\bar\mu}{\bar 
\mu}\gamma_{H_r}(\alpha_s(\bar\mu))\right]
H_r(M_W)\;,
\nonumber \\
H(z,t,\mu)&=&\exp\left[ -\int^{t}_{\mu}\frac{d\bar\mu}{\bar
\mu}\lbrack
\sum_j \gamma_{J_j}(\alpha_s(\bar\mu))+\gamma_S(\alpha_s(\bar\mu))
+\gamma_{H_r}(\alpha_s(\bar\mu))\rbrack\right]
H(z,t)\;,
\label{un}
\end{eqnarray}
$\gamma_{J_j}=-2\alpha_s/\pi$, $\gamma_S=-C_F\alpha_s/\pi$ and
$\gamma_{H_r}$ being the anomalous dimensions of the jet function, the soft
function and the harder function, respectively. $\tilde{J}_j(b)$,
$f(z,b)$ and $H_r(M_W)$ on the right-hand sides are the initial conditions
of the RG evolution, and do not contain large logarithms. Therefore,
$\tilde{J}_j(b)$ are approximated by the tree-level delta functions just
like ${\tilde J}_i(b)$, and $H_r(M_W)$ takes its lowest-order expression
$H_r(M_W)=1$. The $B$ meson distribution function $f(z,b)$ is the same as
that employed in the semileptonic decays.
The scale $t$ should be chosen as the typical scale of the hard
subprocess in order to minimize the involved large logarithms.
A natural choice is the multiple of the maximal relevant scales,
\begin{equation}
t = \kappa\max \left[ {\bar p}_j, \frac{1}{b} \right]\;,
\end{equation}
with $\kappa$ an adjustable parameter. Though other scales besides $t$
still lurk in the hard part, the logarithms they generate are relatively
small. Since $t$ depends on the quark kinematics, it reflects the details
of the meson dynamics. $t$ remains a hard scale as long as the large $b$
region is Sudakov suppressed by $\exp(-s)$, and the perturbative
calculation of the initial condition $H(z,t)$ is reliable. Hence, $H$
assumes its lowest-order expression in Eq.~(\ref{hi}).

Combining the exponents in Eq.~(\ref{un}), the QCD evolution from $M_W$
to $t$, governed by $\gamma_{H_r}$, is identified as the Wilson
coefficients. Conventionally, the summation of the
large logarithms $\ln(M_W/\mu)$ is performed in terms of the effective
Hamiltonian with $W$ boson integrated out. The short-distance physics
related to the $W$ boson are incorporated into Wilson coefficients.
For nonleptonic decays, the relevant effective Hamiltonian (ignoring the
penguin operators for the moment) is
\begin{equation}
H_{\rm eff} = \frac{4G_{F}}{\sqrt{2}} V_{cb} V^{\ast}_{cs}
          [ \, c_{1}(\mu) O_{1} +
            c_{2}(\mu) O_{2} \, ]\;,
\label{eff}
\end{equation}
with the four-fermion current-current operators
\begin{eqnarray}
O_{1} =     (\bar{s}_{L}\gamma_{\mu} b_{L})
                 (\bar{c}_{L}\gamma^{\mu} c_{L})\;,  \;\;\;\;
O_{2} =     (\bar{c}_{L}\gamma_{\mu} b_{L})
                 (\bar{s}_{L}\gamma^{\mu} c_{L})  \;.
\end{eqnarray}
$O_{1,2}$ require renormalization and thus the Wilson coefficients
$c_{1,2}(\mu)$ depend on the renormalization scale $\mu$. It is simpler
to work with the operators $O_{\pm}=\frac{1}{2} (O_{2} \pm O_{1})$ and their
corresponding coefficients $c_{\pm} = c_{2}(\mu) \pm c_{1}(\mu)$, because
they are multiplicatively renormalized.

At last, the nonleptonic decay rates are written as
\begin{eqnarray}
\frac{\Gamma}{\Gamma_0}
&=&\frac{M_B^2}{2}\int\, dx\,dy\,dy_0\,dz
\int_0^{\infty} db\mbox{\ } b \, 
\left[ \frac{N_c\!+\!1}{2}
c_{+}^2(t) + \frac{N_c\!-\!1}{2} c_{-}^2(t) \right] \nonumber \\
&& \times H(z,t)\prod_i {\tilde J}_i(b)\prod_j {\tilde J}_j(b) f(z, b) \, 
\exp\left[-\sum_j s({\bar p}_j,b)-s_{JS}(t,b)\right]
\label{non}
\end{eqnarray}
with
\begin{equation}
s_{JS}(t,b)=\int^{t}_{1/b}\frac{d\bar\mu}{\bar\mu}\left[
\sum_j\gamma_{J_j}(\alpha_s(\bar\mu))+\gamma_S(\alpha_s(\bar\mu))
\right]\;.
\end{equation}
The expressions of the Wilson coefficients $c_\pm$ up to next-to-leading
order are referred to \cite{buras}. Hence, the three-scale factorization
theorem contains the two-stage evolutions: from $M_W$ to $t$ due to the
summation of the logarithms $\ln (M_W/t)$ organized into the Wilson
Coefficients, and from $t$ to $1/b$ due to the summation of the
logarithms $\ln (t b)$ into the exponential $\exp(-s_{JS})$.

We compute the $B$ meson decay rates of the various modes for different
choices of the hard scale parameter $\kappa$. The kinematic inputs are
$M_B=5.28$ GeV, $m_b=4.6$ GeV, which is determined by the first moment
${\bar \Lambda}/M_B=1-m_b/M_B$ of the distribution function in
Eq.~(\ref{bdf}), $M_D=1.87$ GeV, and $m_c=1.6$ GeV. Our predictions
for $B_{\rm SL}$ and $n_c$ are not sensitive to the variation of
$m_c$. The difference between the values for $m_c=1.6$ GeV and for
$m_c=1.5$ GeV is less than 5\%. For the next-to-leading-order Wilson
coefficients, we adopt the HV scheme. The difference between the HV and
NDR schemes has also been examined, and found to be less than 5\%.
In Table I.1 we list the results in the obvious notations $r$, with the
partial decay rates normalized to $B(\bar{B}\to X_c e \bar{\nu})$. This
table is compared to the HQEFT predictions in Table I.2 \cite{neubert}.
On the experimental side, the CLEO group reports
$B_{\rm SL}=10.19 \pm 0.37 \%$, $n_c=1.12 \pm 0.05 \%$ and
recently $B(b\to c \bar{c} s)=21.9\pm 3.7 \%$ \cite{CLEONEW}, which are
only marginally consistent with the LEP measurements
$B_{\rm SL}=11.12 \pm 0.20 \%$ and $n_c=1.20 \pm 0.07 \%$.

Our predictions show a tendency that the experimental data can be
accommodated if the hard scale is chosen properly. Unlike the
next-leading-order HQEFT calculation, we can lower the semileptonic
branching ratio into the range of experiment data with the charm counting
almost remaining fixed. The value of $B(b\to c \bar{c} s)$ in Table I.1 is
consistent with the CLEO measurement. The $b\to c\bar{u} d$ mode gains
larger enhancement than the $b\to c \bar{c} s$ mode in our approach,
because of the larger available phase space for the QCD evolutions.
Therefore, we predict a larger $r_{ud}$ than that from HQEFT. On the
contrary, the calculation of Bagan {\it et al.} shows that the increase of
the mode $b \to c{\bar c}s$ from $O(\alpha_s)$ QCD corrections
dominates \cite{bagan}. Though there is an experimantal argument for
$r_{ud}$ to be about $4.0$ \cite{browder}, the scene is still problematic
\cite{CLEONEW}. Indeed, if $r_{ud}$ is about $4.0$ and $B(b\to c\bar{c}s)$
is $21.9 \%$, which is determined more accurately than $r_{ud}$ by CLEO,
the branching ratios of the various modes will not add up to unity
\cite{CLEONEW}. To render the total branching ratios equal to $100 \%$,
$r_{ud}$ would be $5.2 \pm 0.6$, consistent with our predictions.
Our results of the $B$ meson lifetime $\Gamma$,
whose errors come from the uncertainty
of $|V_{cb}|=0.041\pm 0.005$,  are also consistent with
the data $(1.62\pm 0.06)\times 10^{-12}$ sec.

Note that our calculation leads to $r_{\tau\nu}=0.378$, which is much
larger than the HQEFT result $0.23$. A simple investigation indicates that
$r_{\tau\nu}$ depends on the $B$ meson distrubution function. By varying
the distribution function, a smaller $r_{\tau\nu}$ can be obtained.
However, we stress that the distribution function employed here
is determined from the decay spectrum of $B\to X_s\gamma$. Therefore, we
urge experimental measurements to decide whether the PQCD approach or the
conventional HQEFT approach gives better results. 



\newpage

Table I.1 PQCD predictions of the various decay modes with the notations
$r_{\tau\nu}=B({\bar B}\to X_c\tau{\bar\nu})/B({\bar B}\to X_ce{\bar\nu})$,
$r_{ud}=B({\bar B}\to X_c)/B({\bar B}\to X_ce{\bar\nu})$, and
$r_{cs}=B({\bar B}\to X_{c{\bar c}})/B({\bar B}\to X_ce{\bar\nu})$.

\hspace{0.2in}
\begin{center} 
\begin{tabular}{lccccccc} \hline
$\kappa$ & $ r_{\tau\nu} $ & $r_{ud}$ & $r_{cs}$ 
& $B_{c{\bar c}s}$  & $B_{\rm SL}$ & $n_c$ & $\Gamma$ ($10^{-12}$ sec) \\ 
\hline
$1.0$ & $0.378$ & $4.05$ & $1.55$ & $19.41$ & $12.52$ & $1.195$ 
& $1.50\pm 0.37$ \\ 
$1.5$ & $0.378$ & $4.76$ & $1.74$ & $19.59$ & $11.26$ & $1.196$ 
& $1.35\pm 0.33$\\
$2.0$ & $0.378$ & $5.44$ & $1.88$ & $19.36$ & $10.30$ & $1.194$ 
& $1.24\pm 0.30$ \\
\hline  
\end{tabular}
\end{center}
\hspace{0.2in}

Table I.2 HQEFT predictions of the various decay modes.

\hspace{0.2in}
\begin{center}
\begin{tabular}{lcccccc}
\hline
$\mu/m_b$ & $ r_{\tau\nu} $ & $r_{ud}$ & $r_{cs}$ & 
$B_{c\bar{c}s'}$ & $B_{\rm SL}$ & $n_c$ \\ 
\hline
$1.0$ & $0.22$ & $4.21$ & $1.89$ & $22.68$ & $12.0$ & $1.20$  \\ 
$0.5$ & $0.23$ & $4.75$ & $2.20$ & $23.98$ & $10.9$ & $1.21$  \\
\hline  
\end{tabular}
\end{center}
\hspace{0.2in}


\begin{thebibliography}{99}

\bibitem{CLEO}
P. Drell, Report No. hep-ph/9711020, to appear in the Proceedings of the
18th International Symposium on Lepton-Photon Interactions, Hamburg, 
Germany, July, 1997.
\bibitem{LEP}
M. Feindt, Report No. hep-ph/9802380, to appear in the Proceedings of the 
International Eorophysics Conference on High Energy Physics, 
Jerusalem, Israel, August, 1997.
\bibitem{a6}
G. Altarelli and S. Petrarca, Phys. Lett. B {\bf 261}, 303 (1991).
\bibitem{neubert}
For a review, see, for example, M. Neubert, Report No. hep-ph/9801269.
\bibitem{af}
I.I. Bigi, M.A. Shifman, M.G. Uraltsev, and A.I. Vainshtein, Phys. Rev.
D {\bf 50}, 2234 (1994).
\bibitem{kagan}
A. L. Kagan, Phys. Rev. D {\bf 51}, 6196 (1995); B. Grzadowski and W.S. 
Hou, Phys. Lett. B {\bf 272}, 383 (1992).
\bibitem{grinstein}
B. Grinstein and R.F. Lebed, Phys. Rev. D {\bf 57}, 1366 (1998).
\bibitem{A} 
G. Altarelli, G. Martinelli, S. Petrarca, and F. Rapuano, Phys. Lett. 
B {\bf 382}, 409 (1996).
\bibitem{bagan}
E. Bagan, P. Ball, V. Braun and P. Gosdzinsky, Phys. Lett. B {\bf 342}, 362
(1995); {\it ibid.} Nucl. Phys. {\bf B432}, 3 (1994);
E. Bagan, P. Ball, B. Fiol and P. Gosdzinsky, Phys. Lett. B {\bf 351}, 
546 (1995).
\bibitem{NS}
M. Neubert and C. T. Sachrajda, Nucl. Phys. {\bf B483}, 339 (1997).
\bibitem{sterman}
J. C. Collins, D. E. Soper and G. Sterman, in {\em Perturbative Quantum 
Chromodynamics}, edited by A.H. Mueller (World Scientific, Singapore, 1989). 
\bibitem{LiYu}
H-n. Li and H.L. Yu, Phys. Rev. D {\bf 53}, 4970 (1996).
\bibitem{LiL}
H-n. Li, Phys. Lett. B {\bf 369}, 137 (1996).
\bibitem{N}
M. Neubert, Phys. Rev. D {\bf 49}, 4623 (1994).
\bibitem{btoxgamma}
H-n. Li and H.L. Yu, Phys. Rev. D {\bf 55}, 2833 (1997). 
\bibitem{LiRe}
H-n. Li, Phys. Rev. D {\bf 55}, 105 (1997).


\bibitem{LiCh}
C.H. V. Chang, H-n. Li, Phys. Rev. D {\bf 55}, 5577 (1997).
\bibitem{LiTs}
H-n. Li and B. Tseng, Phys. Rev. D {\bf 57}, 443 (1998). 
\bibitem{BS}
J. Botts and G. Sterman, Nucl. Phys. {B325}, 62 (1989);
H-n. Li and G. Sterman, Nucl. Phys. {B381}, 129 (1992).
\bibitem{buras}
G. Buchalla, A.J. Buras and M.E. Lautenbacher, Rev. Mod. Phys. {\bf 68},
1125 (1996).
\bibitem{CLEONEW}
T. E. Coan et al (CLEO Collaboration), Report No. hep-ph/9710028, to
appear in Phys. Rev. Lett.
\bibitem{browder}
T. E. Browder, Report No. hep-ph/9802217.

\end{thebibliography}
\end{document}